\pdfoutput=1

\documentclass[11pt,twoside,a4paper,cmspaper,final,collab]{cms-tdr}
\def\svnVersion{377933}\def\cmsCernNoTag{CERN-EP/2016-132}\def\cmsCernDate{\today}\def\cmsMessage{Published in the Journal of High Energy Physics as \href{http://dx.doi.org/10.1007/JHEP12(2016)013}{\doi{10.1007/JHEP12(2016)013.}}}

\begin{document}\cmsNoteHeader{SUS-15-011}

\hyphenation{had-ron-i-za-tion}
\hyphenation{cal-or-i-me-ter}
\hyphenation{de-vices}
\RCS$Revision: 372897 $
\RCS$HeadURL: svn+ssh://svn.cern.ch/reps/tdr2/papers/SUS-15-011/trunk/SUS-15-011.tex $
\RCS$Id: SUS-15-011.tex 372897 2016-11-04 16:23:20Z pablom $
\newlength\cmsFigWidth
\ifthenelse{\boolean{cms@external}}{\setlength\cmsFigWidth{0.85\columnwidth}}{\setlength\cmsFigWidth{0.4\textwidth}}
\ifthenelse{\boolean{cms@external}}{\providecommand{\cmsLeft}{top\xspace}}{\providecommand{\cmsLeft}{left\xspace}}
\ifthenelse{\boolean{cms@external}}{\providecommand{\cmsRight}{bottom\xspace}}{\providecommand{\cmsRight}{right\xspace}}
\providecommand{\PV}{\mathrm{V}\xspace}
\newcommand{\nj}{\ensuremath{N_{\text{jets}}}\xspace}
\newcommand{\nb}{\ensuremath{N_{\text{\PQb-jets}}}\xspace}
\newcommand{\mll}{\ensuremath{m_{\ell\ell}}\xspace}
\newcommand{\slep}{\ensuremath{\widetilde{\ell}}\xspace}
\newcommand{\MuMu}{\ensuremath{{\Pgm^\pm\Pgm^\mp}}\xspace}
\newcommand{\ElEl}{\ensuremath{{\Pe^\pm\Pe^\mp}}\xspace}
\newcommand{\EM}{\ensuremath{{\Pe^\pm\Pgm^\mp}}\xspace}
\newcommand{\rmue}{\ensuremath{r_{\PGm/\Pe}}\xspace}
\newcommand{\Rsfof}{\ensuremath{R_{\mathrm{SF/OF}}}\xspace}
\newcommand{\RT}{\ensuremath{R_{\mathrm{T}}}\xspace}
\newcommand{\rinout}{\ensuremath{r_{\text{out/in}}}\xspace}
\newcommand{\vv}{\ensuremath{\PV+\PV}\xspace}
\newcommand{\zjets}{\ensuremath{\Z+\text{jets}}\xspace}
\newcommand{\dyjets}{\ensuremath{\mathrm{DY}+\text{jets}}\xspace}
\newcommand{\gjets}{\ensuremath{\gamma+\text{jets}}\xspace}
\newcommand{\ttz}{\ensuremath{\ttbar\Z}\xspace}
\newcommand{\ttv}{\ensuremath{\ttbar\PV}\xspace}
\newcommand{\njets}{\ensuremath{N_{\text{jets}}}\xspace}
\newcommand{\lint}{2.3\xspace}

\cmsNoteHeader{SUS-15-011}
\title{Search for new physics in final states with two opposite-sign, same-flavor leptons, jets, and missing transverse momentum in pp collisions at \texorpdfstring{$\sqrt{s}=13$\TeV}{sqrt(s)=13 TeV}}

\date{\today}

\abstract{
A search is presented for physics beyond the standard model in final states with two opposite-sign, same-flavor leptons, jets,
and missing transverse momentum. The data sample corresponds to an integrated luminosity of \lint\fbinv of proton-proton collisions
at $\sqrt{s}=13$\TeV collected with the CMS detector at the LHC in 2015. The analysis uses the invariant mass of the lepton pair, searching
for a kinematic edge or a resonant-like excess compatible with the Z boson mass. Both search modes use several event categories in order to increase the sensitivity to
new physics. These categories are based on the rapidity of the leptons, the multiplicity of jets and b jets, the scalar sum of jet transverse momenta,
and missing transverse momentum. The observations in all signal regions are consistent with the expectations from the standard model, and the results
are interpreted in the context of simplified models of supersymmetry.}

\hypersetup{%
pdfauthor={CMS Collaboration},%
pdftitle={Search for new physics in final states with two opposite-sign, same-flavor leptons, jets, and missing transverse momentum in pp collisions at sqrt(s) = 13 TeV},%
pdfsubject={CMS},%
pdfkeywords={CMS, physics, supersymmetry}}

\maketitle

\clearpage

\section{Introduction}
\label{sec:introduction}

Supersymmetry (SUSY)~\cite{Ramond:1971gb,Golfand:1971iw,Neveu:1971rx,Volkov:1972jx,Wess:1973kz,Wess:1974tw,Fayet:1974pd,Nilles:1983ge}
is one of the most appealing extensions of the standard model (SM), assuming a new fundamental
symmetry that assigns a new fermion (boson) to every SM boson (fermion). SUSY resolves the hierarchy problem of the SM
by stabilizing the Higgs boson mass via additional quantum loop corrections from the top super-partner (top squark), which compensate the correction due to the top quark.
If $R$-parity~\cite{Farrar:1978xj} is conserved the lightest state predicted by the theory is stable and potentially massive, providing a candidate for Dark Matter. Many SUSY models also
lead to the unification of the electroweak and strong forces at high energies.

This paper presents a search for signatures of SUSY in events with two opposite-sign, same-flavor leptons (electrons or muons), jets, and missing transverse momentum. A
dataset of pp collisions collected with the CMS detector at the CERN LHC at a center-of-mass energy $\sqrt{s}=13$\TeV in 2015 was used, corresponding
to an integrated luminosity of \lint\fbinv. The dilepton topology is expected to occur in SUSY models where a neutralino decays to either an on-shell Z boson or a
virtual $\Z/\gamma$ boson which in turn decays to leptons and the lightest SUSY particle (LSP), or into a lepton and its supersymmetric partner (slepton), the latter decaying into another
lepton and the LSP. Decays involving an on-shell Z boson are expected to produce an excess of events
compatible with the Z boson mass, while decays involving off-shell Z bosons or sleptons are expected to produce a characteristic edge shape in the invariant mass distribution of
the dilepton system~\cite{Hinchliffe:1996iu}.

The CMS Collaboration published a version of this analysis using a $\sqrt{s}= 8$\TeV dataset, observing a $2.6\,\sigma$ local significance excess compatible with an edge shape located at a
dilepton invariant mass of $78.7 \pm 1.4$\GeV~\cite{CMS:edge}. The ATLAS collaboration reported the absence of any excess in a similar signal
region, but observed a $3.0\,\sigma$ excess in dilepton events compatible with the Z boson mass~\cite{ATLAS:edge}. Both of these excesses
warrant scrutiny using the 13\TeV dataset and are analyzed here with minor changes with respect to the 8\TeV searches.

\section{The CMS detector}
\label{sec:cmsdetector}

The central feature of the CMS apparatus is a superconducting solenoid, 13~m in length and 6~m in diameter, that provides
an axial magnetic field of 3.8\unit{T}. The bore of the solenoid is outfitted with various particle detection systems. Charged-particle
trajectories are measured by silicon pixel and strip trackers, covering $0 < \phi < 2\pi$ in azimuth and $\abs{\eta} < 2.5$,
where the pseudorapidity $\eta$ is defined as $\eta = -\log [\tan(\theta/2)]$, with $\theta$ being the polar angle of the
trajectory of the particle with respect to the beam direction. A crystal electromagnetic calorimeter (ECAL), and a brass and scintillator
hadron calorimeter surround the tracking volume. The calorimetry provides high resolution energy
and direction measurements of electrons and hadronic jets. A preshower detector consisting of two planes of silicon sensors
interleaved with lead is located in front of the ECAL at $\abs{\eta}>1.479$. Muons are measured in gas-ionization detectors embedded in
the steel flux-return yoke outside the solenoid. The detector is nearly hermetic, allowing for energy balance measurements in the
plane transverse to the beam direction. A two-tier trigger system selects the most interesting pp collision events for use
in physics analysis. A more detailed description of the CMS detector, its coordinate system, and the main kinematic
variables used in the analysis can be found elsewhere~\cite{Chatrchyan:2008zzk}.

\section{Datasets, triggers, and object selection}
\label{sec:samplesObjects}

Events are collected with a set of isolated dilepton triggers that require a transverse momentum $\pt > 17$\GeV for the leading lepton and $\pt > 12\,(8)$\GeV
for the subleading electron (muon), and $\abs{\eta}<2.5\,(2.4)$ for electrons (muons). In order to retain high signal efficiency, in particular for Lorentz-boosted dilepton systems,
non-isolated dilepton triggers with $\pt > 33\,(27)$\GeV for the first electron (muon) and $\pt > 33\,(8)$\GeV for
the second electron (muon) are also used. The trigger efficiencies are measured in data using events selected by a suite of jet triggers.

Events are selected by requiring two opposite-charge, same flavor leptons (\ElEl or \MuMu) with $\pt > 20$\GeV and pseudorapidity $\abs{\eta} <$ 2.4. The distance between the
leptons is requested to be at least $\sqrt{\smash[b]{\Delta\phi^2 + \Delta \eta^2}} = \Delta R > 0.3$ to avoid reconstruction efficiency differences between electrons and muons in events with very
collinear leptons. This requirement is relaxed to $\Delta R > 0.1$ when the mass of the dilepton system is consistent with a Z boson to preserve acceptance for Z bosons
with large transverse momentum. To ensure symmetry in acceptance between electrons and muons, all events with
one of these two leptons in the barrel-endcap transition region of the ECAL, $1.4 < \abs{\eta} <  1.6$, are rejected. A control sample of different
flavor leptons ($\Pe\Pgm$ or $\Pgm\Pe$) is defined using the same lepton selection criteria. All the parameters above have been chosen in order to maximize the lepton selection efficiency while
keeping the electron and muon efficiencies similar.

Electrons, reconstructed by associating tracks with ECAL clusters, are identified using a multivariate approach based on information on the
cluster shape in the ECAL, track quality, and the matching between the track and the ECAL cluster~\cite{Khachatryan:2015hwa}. Additionally, electrons from photon conversions
are rejected. Muons are reconstructed from tracks found in the muon system associated with tracks in the tracker. They are identified
based on the quality of the track fit and the number of associated hits in the tracking detectors. For both lepton flavors, the impact
parameter with respect to the reconstructed vertex with the largest $\pt^2$ sum of associated tracks (primary vertex) is required to be within 0.5\unit{mm} in the transverse
plane and below 1\unit{mm} along the beam direction. The lepton isolation, defined as the scalar \pt sum of all particle candidates, excluding the lepton itself,
in a cone around the lepton, divided by the lepton \pt, is required to be smaller than 0.1\,(0.2) for electrons (muons). A cone-size, varying with lepton \pt, is chosen to be
$\Delta R = 0.2$ for $\pt < 50$\GeV, $\Delta R = 10\GeV/\pt$ for $50 < \pt < 200\GeV$, and $\Delta R = 0.05$ for $\pt > 200$\GeV.

A particle flow (PF) technique~\cite{CMS:2009nxa,CMS:2010byl} is used to reconstruct particle candidates in the event. Jets are clustered from
these candidates, excluding charged hadrons not associated to the primary vertex, using the anti-\kt clustering algorithm~\cite{Cacciari:2008gp}
implemented in the \FASTJET package~\cite{FastJet,Cacciari:2005hq} with a distance parameter of 0.4. Each jet is required to have $\pt > 35$\GeV
where the \pt is corrected for non-uniform detector response and multiple collision (pileup) effects~\cite{1748-0221-6-11-P11002,cacciari-2008-659}, and $\abs{\eta} < 2.4$. A jet is removed
from the event if it lies within $\Delta R < 0.4$ of any of the selected leptons. The scalar sum of all jet transverse momenta is referred to as \HT.
The magnitude of the negative vector \pt sum of all the PF candidates is referred to as \MET. Corrections to the jet energy are propagated to the \MET
using the procedure developed for 7\TeV data~\cite{1748-0221-6-11-P11002}.
Identification of jets originating from b-quarks is performed with the combined secondary vertex algorithm, using a working point in which the typical
efficiency for b quarks is around 65\% and the mistagging rate for light-flavor jets is around 1.5\%~\cite{CMS:2016kkf}.

While the main SM backgrounds are estimated using data control samples, simulated events are used to estimate uncertainties and
minor SM background components. Next-to-leading order (NLO) and next-to-NLO cross sections~\cite{Alwall:2014hca,Alioli:2009je,Re:2010bp,Gavin:2010az,Gavin:2012sy,Czakon:2011xx}
are used to normalize the simulated background samples, while NLO plus next-to-leading-logarithm (NLL) calculations~\cite{Borschensky:2014cia} are used for the signal samples. Simulated samples
of Drell--Yan (DY) production associated with jets (\dyjets), \gjets, \vv, and \ttv ($\PV=\PW,\Z$) events
are generated with the \MADGRAPH\MCATNLO v2.2.2 event generator~\cite{Alwall:2014hca}, while \POWHEG v1~\cite{powheg} is used for \ttbar and
single top quark production. The matrix element calculations performed with these generators are interfaced with \PYTHIA8~\cite{Sjostrand:2007gs} for the
simulation of parton showering and hadronization. The NNPDF3.0 parton distribution functions (PDF)~\cite{Ball:2014uwa} are used for all samples.
The detector response is simulated with a \GEANTfour model~\cite{Geant} of the CMS detector.
The simulation of new physics signals is performed using the \textsc{MadGraph5\_aMC@NLO}\xspace program at LO precision with up to
2 additional partons in the matrix elements calculations. Events are then interfaced with \PYTHIA~8 for fragmentation and hadronization, and simulated using the CMS fast simulation package~\cite{fastsim}. Multiple pp interactions are superimposed on the hard
collision and the simulated samples are reweighted to reflect the beam conditions. Normalization scale factors are applied to the simulated samples to
account for differences between simulation and data in the trigger and reconstruction efficiencies.

\section{Signal models}
\label{sec:signalmodels}

This search targets different modes of neutralino decays into final
states with two opposite-sign, same-flavor leptons, jets, and \ETm originating from the LSPs. In order
to study these processes, two simplified models have been considered for the two search modes: one producing a resonant
lepton signature through an on-shell Z boson for the ``on-Z'' search, and another producing an edge-like distribution in the invariant mass of the leptons, for
the ``edge'' search.

\begin{figure}
\centering
\includegraphics[width=0.4\textwidth]{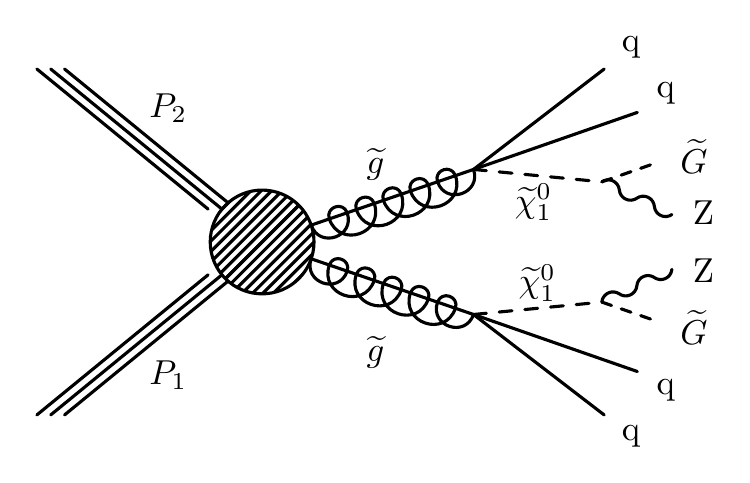}
\includegraphics[width=0.4\textwidth]{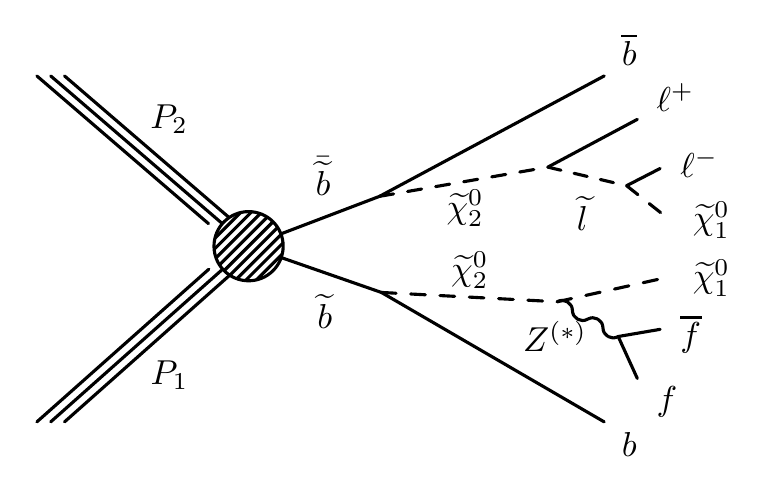}
\caption{Diagrams for gluino and \PSQb pair production and decays realized in the simplified models.
The GMSB model targeted by the on-Z search is shown on the left. On the right, the slepton-edge model features characteristic edges in the \mll spectrum given by the mass difference of the \PSGczDt and \PSGczDo.}
\label{sig:feynmanEdge}
\end{figure}

The first of these simplified models represents gauge mediated supersymmetry breaking SUSY models~\cite{Martin:1997ns} and
is referred to as the GMSB scenario. The model assumes the production of a pair
of gluinos (\PSg) that decay into a pair of quarks (u, d, s, c, or b) and the lightest neutralino \PSGczDo. This
neutralino decays into an on-shell Z boson and a massless gravitino (\PXXSG) as seen in Fig.~\ref{sig:feynmanEdge} (left).
At least one of the Z bosons decays into a pair of leptons producing the signature targeted by the on-Z search.

The signal model for the edge search, referred to as slepton-edge, assumes
the production of a pair of bottom squarks, which decay to the next-to-lightest neutralino \PSGczDt
and a b-quark. Two decay modes of the \PSGczDt are considered each with 50\% probability. In the first one,
the \PSGczDt decays to a Z boson and the lightest neutralino \PSGczDo, which is stable. The Z boson
can be on or off-shell, depending on the mass difference between the neutralinos, and decays according
to its SM branching fractions. The second one features subsequent two-body decays with an intermediate
slepton \slep: $\PSGczDt \to\slep \ell \to\ell\ell\PSGczDo$. The masses
of the sleptons ($\PSe$,$\PSgm$) are assumed degenerate and equal to the average of the \PSGczDt and \PSGczDo.
The masses of the \PSQb and \PSGczDt are free parameters, while $m_{\PSGczDo}$ is fixed at 100\GeV. This scheme allows the
position of the signal edge to vary along the invariant mass distribution according to the mass difference between the \PSGczDt and \PSGczDo. The mass
of the \PSGczDo has been chosen in such a way that the difference to the \PSGczDt mass is above 50\GeV, setting the minimum possible edge position at 50\GeV. An
example for one of the possible decays is shown in Fig.~\ref{sig:feynmanEdge} (right).

\section{Signal regions}
\label{subsec:signalregions}

Signal regions for the on-Z and edge searches follow two principles: first, they are designed to provide sensitivity
to a range of new physics models, including the simplified models defined above, and second, they are designed to
investigate excesses in the 8\TeV datasets reported by the ATLAS and CMS Collaborations~\cite{ATLAS:edge,CMS:edge}.
The selections described below are applied in addition to the dilepton selection described in Section~\ref{sec:samplesObjects}.

\subsection{On-Z signal regions}
\label{subsub:onzSR}

The on-Z search is divided into a total of three signal region (SR) categories with dilepton invariant mass (\mll) in the range $81 < \mll < 101$\GeV. The first two, referred to as
``SRA'' (2--3 jets and $\HT > 40$0\GeV) and ``SRB'' ($\geq$4 jets), focus on events with low and high jet multiplicity. These categories are further
divided according to the number of b-tagged jets and \MET. One additional signal region, namely ``ATLAS SR'', is defined corresponding to the
region showing a $3.0\,\sigma$ excess in the 8\TeV dataset of the ATLAS Collaboration~\cite{ATLAS:edge}. The selection details are specified in Section~\ref{sec:results}.

\subsection{Edge search signal regions}
\label{subsub:offzSR}

The signal regions in the edge search remain largely unchanged with respect to the search performed
with the 8\TeV dataset~\cite{CMS:edge}. The requirements on the jet multiplicity and
\MET are similar to the previous analysis, namely $\MET > 100\,(150)$\GeV if at least three (two) jets
are present. The relative centrality
expected in the decays of heavy particles, combined with the performance of the detector
in the barrel region compared to the endcaps, motivates a division of the event sample depending
on the $\abs{\eta}$ of the leptons. The signal region is defined as central
if both leptons lie within $\abs{\eta}<1.4$ and as forward if at least one of the leptons is
located outside of this $\abs{\eta}$ range. Furthermore, two exclusive bins are defined in the number of b-tagged
jets, one without and one with at least one such jet.

The improvements in the CMS reconstruction algorithms for the 13\TeV data taking lead to a few
differences between the 8 and 13\TeV signal regions. The lepton identification algorithms have been
updated for the 13\TeV data taking, with the most relevant improvement being the use of a new electron identification
algorithm based on a multivariate discriminator~\cite{Khachatryan:2015hwa}. The jet momentum threshold has been lowered from
40\GeV to 35\GeV given the improved pile-up rejection achieved at $\sqrt{s}= 13$\TeV, and the maximum $\abs{\eta}$
has been reduced from to 3.0 to 2.4, to match the tracker acceptance. The isolation definition has
also been modified to include a variable cone size. The rejection of non-prompt leptons has been improved as a consequence of all these changes. Finally, additional
non-isolated double-lepton triggers have been added to recover efficiency for very boosted dilepton systems,
although the increase in efficiency for the edge signal regions has been found to be small ($<$4\%).

A counting experiment is performed in five distinct regions of the \mll spectrum with events split among
the four exclusive (0 or $>=$1 b-tagged jet, central or forward) and two inclusive (central or forward) categories. The
five mass regions include the three that were present in the 8\TeV analysis (the low-mass region: $20 < \mll < 70$\GeV,
the on-Z region: $81 < \mll < 101$\GeV, and the high-mass region: $\mll > 120$\GeV), as well as the two regions immediately
adjacent to the Z peak ($70 < \mll < 81$\GeV and $101 < \mll < 120$\GeV). The mass spectrum in the current analysis
thus covers all \mll values above 20\GeV.

In order to directly compare the result obtained at 13\TeV with those obtained at 8\TeV, results for
the signal regions are also given inclusively in the number of b-tagged jets, $\nb \geq 0$. A summary of all
signal regions is given along with the experimental results in Section~\ref{sec:results}.

\section{Standard model background predictions}
\label{sec:backgrounds}

The backgrounds from SM processes are divided into two types. Those that produce opposite-flavor (OF) pairs (\EM)
as often as same-flavor (SF) pairs (\MuMu, \ElEl) are referred to as flavor-symmetric (FS) backgrounds. Among them, the
dominant contribution arises from top quark-antitop quark production; sub-leading contributions from WW, $\Z/\gamma^{*} (\to\tau\tau)$,
tW single-top quark production, and leptons from hadron decays are also present. The other category of backgrounds includes flavor-correlated lepton production
and only contributes with SF leptons. The dominant contributions arise from DY production in association with jets, where the \MET arises from mismeasurement of the jet energies. Smaller contributions come from WZ and ZZ production, as well as rare processes such as \ttz. These backgrounds are referred to as ``Other SM'' in this paper.

\subsection{Flavor-symmetric backgrounds}
\label{sub:fsbkg}
The contribution of flavor-symmetric processes in the SF channels is estimated from the OF control sample.
While there is a production symmetry between the two channels at particle level, it can be distorted by the different
trigger, reconstruction, and identification efficiencies for electrons and muons. The background estimate
is therefore obtained from the observed OF yield by applying a multiplicative correction factor, \Rsfof.
This factor is determined by two independent methods, a direct measurement in a control region enriched in
FS backgrounds, and from the measurement of lepton efficiencies, factorized into the effects of reconstruction, identification,
and trigger.

The direct measurement is performed in the region with $\njets = 2$ and $100 < \MET <  150$\GeV, excluding the
mass range $70 < \mll < 110$\GeV to reduce background contributions from resonant Z-boson production. Here, \Rsfof is
evaluated using the observed yield of SF and OF events, $4\Rsfof = N_\mathrm{SF}/N_\mathrm{OF}$. The applicability of this value
in the signal region is confirmed by comparing it with the \Rsfof value obtained in the signal region for \ttbar
simulated events. The difference between both values is found to be smaller than its statistical uncertainty (3\%). The latter
value is assigned as the systematic uncertainty in the measurement.

For the factorized approach, the ratio of muon to electron reconstruction and identification efficiencies, \rmue,
is measured in a DY-enriched region with $\njets \geq 2$ and $\MET < 50$\GeV and requiring $60 < \mll < 120$\GeV,
resulting in a large sample of \ElEl and \MuMu events with similar kinematics to the signal region in terms of jet multiplicity.
Assuming the factorization of lepton efficiencies in an event, the efficiency ratio is measured as
$\rmue = \sqrt{\smash[b]{{N}_{\PGmp\PGmm}/{N}_{\Pep\Pem}}}$. A systematic uncertainty of 10\%\,(20\%) is assigned to \rmue
in the central (forward) lepton rapidity selection based on studies of its dependency on the lepton kinematics, the amount of \MET, and the jet multiplicity. The trigger efficiencies
for the three different flavor combinations are used to define the factor \RT = $\sqrt{\smash[b]{\epsilon^\mathrm{T}_{\MuMu}\epsilon^\mathrm{T}_{\ElEl}}}/\epsilon^\mathrm{T}_{\Pe^{\pm}\PGm^{\mp}}$,
which takes into account the difference between SF and OF channels at the trigger level. The final correction is \Rsfof = $(1/2)(r_{\PGm/\Pe}+r_{\PGm/\Pe}^{-1})\, R_\mathrm{T}$.
Here, \rmue is summed with its inverse, leading to a large reduction of the associated uncertainty.

The results of the direct measurement and the factorization method are shown in Table~\ref{tab:combinedRSFOF}.
Since the results are in agreement and are obtained on independent data samples, they are combined using the
weighted average. The resulting correction is \Rsfof = $1.03\pm0.05$ ($1.08\pm0.07$) for the
central (forward) lepton rapidity selection.

\begin{table}[hbtp]
\centering
\topcaption{Summary of $R_\mathrm{SF/OF}$ values obtained in data and simulation using the direct and factorized methods, and the final combination.}
\label{tab:combinedRSFOF}
\begin{tabular}{l c c c c }
\hline
\multicolumn{1}{c}{} & \multicolumn{2}{c}{{Central}} & \multicolumn{2}{c}{{Forward}} \\ \hline
& Data & MC & Data & MC \\ \hline
$(1/2)( \rmue + \rmue^{-1} )$   &  1.01 $\pm$ 0.01  &   1.01 $\pm$ 0.01   &   1.02 $\pm$ 0.04 &    1.03 $\pm$ 0.05    \\
$R_{T}$                          &  1.00 $\pm$ 0.07  &   1.02 $\pm$ 0.06   &   1.04 $\pm$ 0.09 &    1.04 $\pm$ 0.06    \\ \hline
\Rsfof & & & &  \\
From factorization        &  1.01 $\pm$ 0.07   &  1.03 $\pm$ 0.06       &  1.06 $\pm$ 0.10  &   1.05 $\pm$ 0.08     \\
     Direct measurement   &  1.05 $\pm$ 0.06   &  1.05 $\pm$ 0.03       &  1.10 $\pm$ 0.09  &   1.08 $\pm$ 0.04     \\ \hline
Weighted average    &  {1.03 $\pm$ 0.05}   &  {1.04 $\pm$ 0.03}       &  {1.08 $\pm$ 0.07}  &   {1.07 $\pm$ 0.04}     \\
\hline
\end{tabular}
\end{table}

\subsection{Drell--Yan-like backgrounds}
\label{sub:dybkg}

The \MET from the DY background is estimated from \MET templates obtained from a data control region. The main premise of this estimate based on data is
that \MET in \zjets events originates from the limited detector resolution when measuring the objects making up the hadronic system that recoils against the Z boson.
We estimate the shape of the \MET distribution from a control sample of \gjets events where the jet system recoils against a photon instead of a Z boson.
Signal regions requiring at least one b-tagged jet can lead to a small amount of additional \MET due to the neutrinos in semileptonic b quark decays.
To account for this effect, the \MET templates are extracted from a control sample of \gjets events with at least one b-tagged jet.

The $\gamma +\text{jets}$ events in data are selected with a suite of single-photon triggers with \pt thresholds varying from 22 to 165\GeV.
The triggers with thresholds below 165\GeV are prescaled such that only a fraction of accepted events are recorded,
and the events are weighted by the trigger prescales to match the integrated luminosity collected with the signal dilepton triggers.
In order to account for kinematic differences between the hadronic systems in the \gjets and the \zjets sample,
the \gjets sample is reweighted such that the boson \pt distribution matches that of the \zjets sample.
This reweighting is performed for each signal region, where the same requirements are applied to the \zjets and the \gjets samples.
The resulting \MET distribution is then normalized to the observed data yield in the region $\MET < 50$\GeV where \zjets is the dominant background.

The control sample used to estimate this background does not need to have a high purity of photons, since the \MET is assumed to originate from jet mismeasurement.
However, it is required that the photon-like object be well measured so as to not contribute to the \MET mismeasurement. The stability of the photon selection is tested
by repeating this background measurement after tightening the photon ID requirements, and it is found that the results are consistent with the measurement done using
the looser selection. In order to ensure the photon-like object is sufficiently well-measured and that the \MET in the \gjets sample comes primarily from the mismeasurement of the jet system,
the following conditions are required: $\Delta\phi(\MET, \gamma) > 0.4$, a veto on events where the photon can be connected to a pattern of hits in the pixel
detector, and the photon to be matched to a jet within a cone of $\Delta R = 0.4$.
The requirement $\Delta\phi(\MET, \gamma) >0.4$ protects against under-measurement of the photon energy,
which is much more likely for calorimeter-based quantities than over-measurement.
Finally, the electromagnetic fraction of the matched jet (fraction of jet energy deposited in the electromagnetic calorimeter
with respect to the total energy deposited in both, the electromagnetic and hadronic calorimeter) is required to be $>$0.7.

The dominant uncertainties in the \MET template prediction come from the limited size of the samples used.
The uncertainty in the prediction takes into account the statistical uncertainty of the \gjets sample in the signal \MET regions, which ranges from 10--50\%.
The statistical uncertainty of the normalization for $\MET < 50$\GeV is included and ranges from 4--10\%,
as shown in Table~\ref{tab:template_systematics_reweighting}.
A closure test of the method is performed in simulation, using \gjets to predict the yield of \zjets.  An uncertainty is assigned from
the results of this test as either the largest discrepancy between the \gjets prediction and the \zjets yield for each \MET region,
or the MC statistical uncertainty, whichever is larger. The values are listed in Table~\ref{tab:template_systematics_overview}
and vary between 4 and 50\%, depending on the \MET region.
Finally, the impact of photon purity on the estimate is studied in data by repeating the prediction with
a tighter photon selection. Since the difference from the nominal prediction was smaller than the statistical
uncertainty in all regions, no additional uncertainty was assigned.

\begin{table}[htb]
    \topcaption{\label{tab:template_systematics_reweighting}
      Statistical uncertainties in the normalization of the \MET\ template prediction in the $\MET\ <50$\GeV range, for each signal region.
      These are taken as a systematic uncertainty in the background prediction. The definitions of SRA, SRB, and ATLAS SR are found in Section~\ref{subsub:onzSR} and
      Table~\ref{tab:resultsOnZ}.
    }
    \begin{tabular}{l c c c c c}
      \hline
      Signal region & \multicolumn{2}{c}{SRA} & \multicolumn{2}{c}{SRB} & \multicolumn{1}{c}{ATLAS SR}     \\ \hline
      b tagging & b-jet veto & $\geq$ 1 b tag & b-jet veto & $\geq$ 1 b tag & ---  \\
      \hline
      Uncertainty   &        4 \% &               10 \% &        3 \% &                6 \% &                3 \% \\
      \hline
    \end{tabular}

\end{table}

\begin{table}[htb]
\centering
\topcaption{\label{tab:template_systematics_overview}
Systematic uncertainties in percentage for the \MET\ template method from the MC closure test, shown for all the on-Z signal regions. The definitions of SRA, SRB, and ATLAS SR are found in Section~\ref{subsub:onzSR} and Table~\ref{tab:resultsOnZ}.
}
\begin{tabular}{l c c c c c c}
\hline
\ETm\,(\GeVns{}) &0--50 & 50--100 & 100--150  &150--225 & 225--300 & $\geq$ 300 \\
\hline
SRA, b-jet veto             & 1     & 4        &  4         &  5       & 15        & 35         \\
SRA, $\geq$1 b tag         & 1     & 3        &  5         & 10       & 30        & 40         \\
\hline
SRB, b-jet veto             & 1     & 2        &  4         & 10       & 20        & 25         \\
SRB, $\geq$1 b tag         & 2     & 3        & 10         & 10       & 50        & 50         \\
\hline
ATLAS SR                    & 2     & 2        & 10         & 10       & \multicolumn{2}{c}{10} \\
\hline
\end{tabular}
\end{table}

\subsubsection{Other standard model processes with a Z boson}

The method using \MET templates only predicts instrumental \MET from jet mismeasurement
and thus does not include the genuine \MET\ from neutrinos expected in processes like $\PW(\ell\nu)\Z(\ell\ell)$,
$\Z(\ell\ell)\Z(\nu\nu)$, or rarer processes such as \ttz. These processes contribute a small fraction of the overall background and are determined with MC simulation.
The MC prediction is compared to data in 3- and 4-lepton control regions. Agreement is observed, and a conservative uncertainty of 50\% is assigned
based on the limited statistics of these regions at higher jet multiplicities.

\subsubsection{Drell--Yan background in the edge search}

A procedure was designed to propagate the estimations obtained using the \MET templates for the on-Z regions to the off-Z mass regions.
For this reason, a ratio \rinout is measured in the DY-dominated
control region where \rmue is also obtained. The numerator of this ratio is the number of SF events outside of the Z boson
mass window, while the denominator is the SF yield within this window. Opposite-flavor
yields in both the numerator and denominator are subtracted from the respective same-flavor yields in order to correct
for FS contributions in the region where \rinout is measured. The final ratio is unity for the mass region between 81 and 101\GeV,
and varies between 2\% and 7\% for the other mass ranges, with values decreasing as a function of the invariant mass.
The final contribution to the edge-like signal regions is then the on-Z prediction multiplied by this ratio for each
of the signal regions.  An uncertainty of 25\% is assigned to \rinout to cover its dependencies on \MET and the jet multiplicity.

\section{Results}
\label{sec:results}

\begin{figure}[!h]
\centering
    \includegraphics[width=0.6\textwidth]{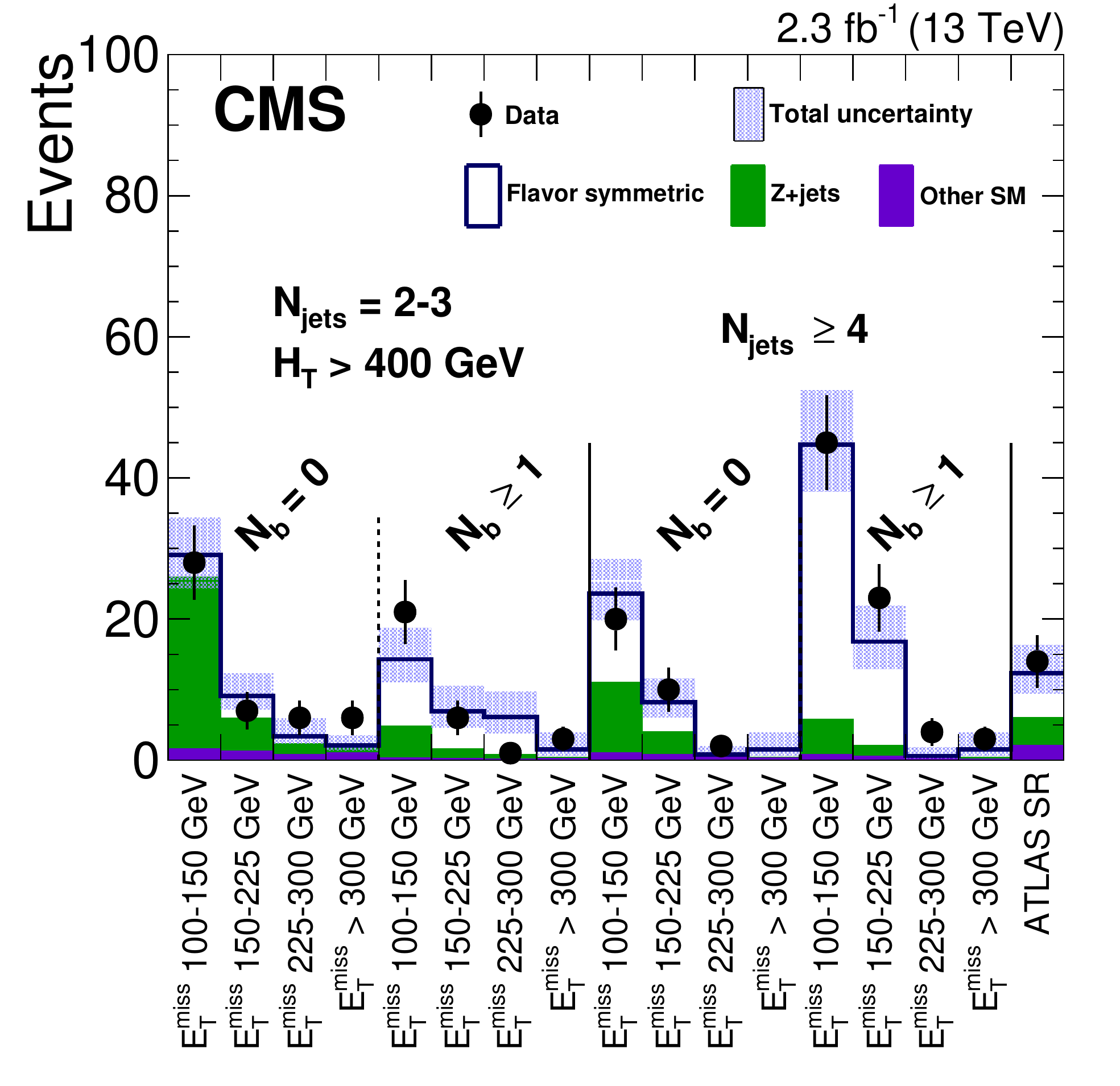}
    \includegraphics[width=0.6\textwidth]{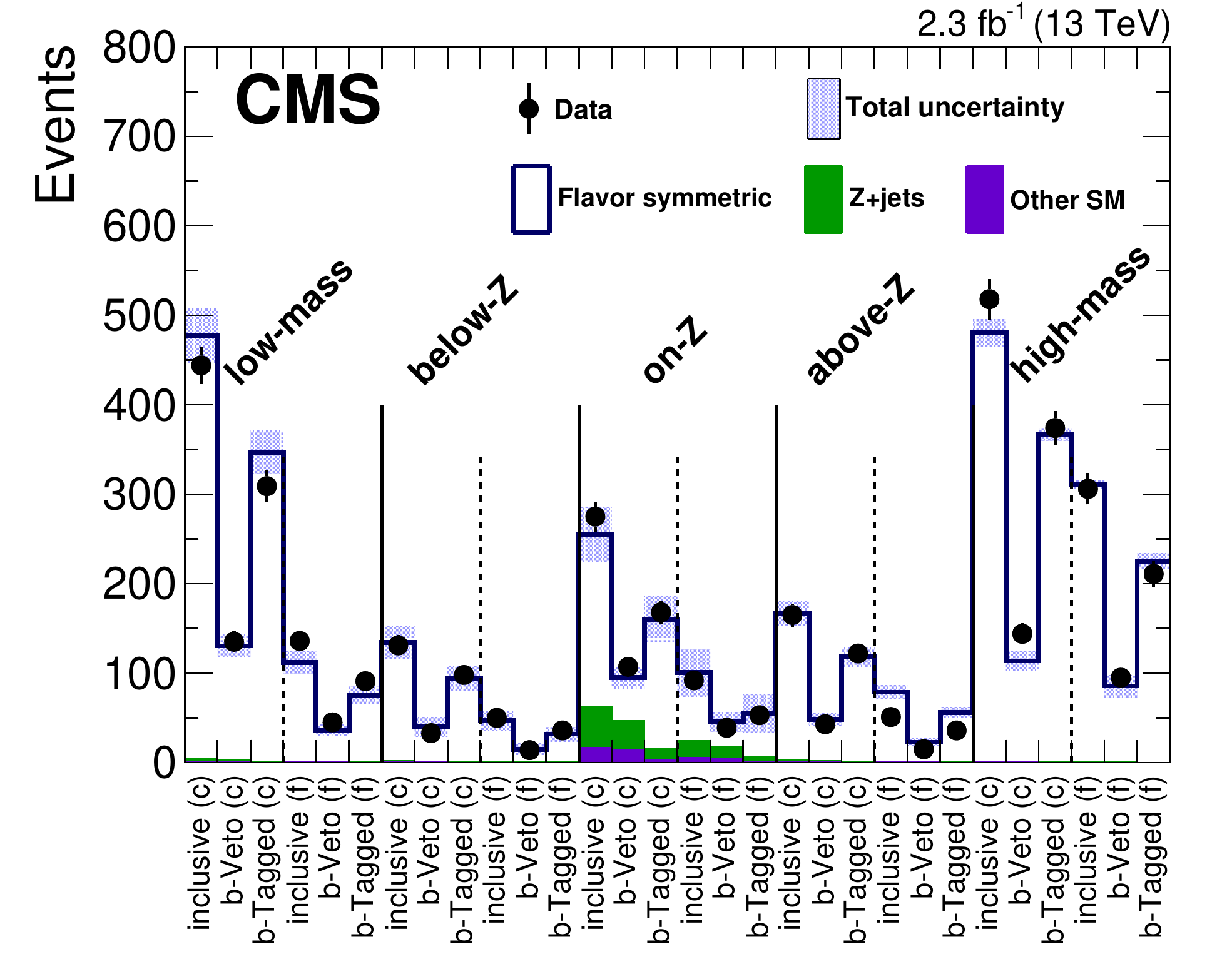}
    \caption{Overview of the results in all signal regions of the on-Z search (upper) and edge search (lower). The labels (c) and (f) refer to central and forward leptons. The data points in black are compared to the background expectation, which is shown as a solid blue line, together with its uncertainty, shown as a light blue band. The background components are shown as a stacked histogram with solid white color for the FS background, solid dark green for DY and dark purple for others.}
\label{fig:onZResultsFancy}
\end{figure}

The observed number of events in the different signal regions is compared with the
background estimates obtained with the methods explained above for the
on-Z and the edge searches. The results for the 16 exclusive signal regions of the on-Z search and the additional
ATLAS signal region are presented in Table~\ref{tab:resultsOnZ}. A graphical representation of these results can be
seen in Fig.~\ref{fig:onZResultsFancy} (upper), where the background prediction has been divided
into its three components: FS, DY, and other processes with a Z boson, in order to illustrate their relative
contributions in the different signal regions.

\begin{table}[!htb]
\renewcommand{\arraystretch}{1.3}
\centering
\topcaption{\label{tab:resultsOnZ}
Observed and predicted yields for the on-Z search.
The signal regions SRA and SRB are binned as a function of the b jet multiplicity and the missing transverse momentum.
In the ATLAS SR, the transverse momenta of the two highest \pt\ leptons are included when calculating \HT,
and an additional requirement is imposed on the angle between the \MET and the two leading jets $\Delta\phi_{\MET,j_1,j_2}>0.4$.}
\begin{tabular}{c | c c c c }
    \hline
                 \nj / \HT                    &  \nb                     & \MET\,(\GeVns)    & Predicted & Observed      \\ \hline
                                            &  \multirow{4}{*}{ 0 }    & 100--150  &  29.1 $^{+5.3}_{-4.7}$ & { 28}  \\
     {SRA}                           &                             & 150--225  &   9.1 $^{+3.2}_{-1.9}$ & {  7}  \\
                                            &                             & 225--300  &   3.4 $^{+2.5}_{-1.0}$ & {  6}  \\
 2--3 jets                                  &                             & $>$300  &   2.1 $^{+1.4}_{-0.7}$ & {  6}  \\ \cline{2-5}
\multirow{4}{*}{and $\HT > 400$\GeV} &  \multirow{4}{*}{$\geq$1 } & 100--150  &  14.3 $^{+4.4}_{-3.2}$ & { 21}  \\
                                            &                             & 150--225  &   6.9 $^{+3.6}_{-2.3}$ & {  6}  \\
                                            &                             & 225--300  &   6.1 $^{+3.6}_{-2.3}$ & {  1}  \\
                                            &                             & $>$300  &   1.5 $^{+2.4}_{-0.9}$ & {  3}  \\ \hline
                                            &  \multirow{4}{*}{ 0 }    & 100--150  &  23.6 $^{+4.9}_{-3.7}$ & { 20}  \\
    {SRB}                            &                             & 150--225  &   8.2 $^{+3.4}_{-2.1}$ & { 10}  \\
                                            &                             & 225--300  &   0.8 $^{+1.2}_{-0.2}$ & {  2}  \\
\multirow{5}{*}{ $\geq$ 4 jets         }    &                             & $>$300  &   1.5 $^{+2.4}_{-0.9}$ & {  0}  \\ \cline{2-5}
                                            &  \multirow{4}{*}{$\geq$1 } & 100--150  &  44.7 $^{+7.7}_{-6.6}$ & { 45}  \\
                                            &                             & 150--225  &  16.8 $^{+5.1}_{-3.9}$ & { 23}  \\
                                            &                             & 225--300  &   0.6 $^{+1.2}_{-0.3}$ & {  4}  \\
                                            &                             & $>$300  &   1.5 $^{+2.4}_{-0.9}$ & {  3}  \\ \hline
\multicolumn{5}{c}{ {ATLAS--SR:} }     \\ \hline
\multicolumn{1}{c}{$\HT+\pt^{\ell_{1}}+\pt^{\ell_{2}}>600$\GeV}&  \multicolumn{1}{c}{$\MET > 225$\GeV} & \multicolumn{1}{c}{$\Delta\phi_{\MET,j_1,j_2}>0.4$} & \multicolumn{1}{c}{$12.3^{+4.0}_{-2.8}$} & \multicolumn{1}{c}{{14}} \\
\hline
\end{tabular}
\end{table}

The edge-like search features two distinct \mll spectra according to the centrality of the leptons, each of
which is divided into five bins. This leads to a total of 10 mutually exclusive signal regions that are further
divided according to the presence or absence of any b-tagged jet in the event. To be consistent with the 8\TeV search,
the information without any selection on the number of b-tagged jets is also provided. Table~\ref{tab:edgeResultsTable}
summarizes the SM predictions and the observations in all these signal regions. A graphical representation of these results
is shown in Fig.~\ref{fig:onZResultsFancy} (lower), including the relative contributions of the different backgrounds.

\begin{table}[htb]
\renewcommand{\arraystretch}{1.1}
\centering
\topcaption{\label{tab:edgeResultsTable}
Results for the edge-like search in all 30 signal regions. The non-FS component of the total background is given
separately in the brackets. All signal regions require \MET $>$150\,(100)\GeV if \nj $\geq$ 2 (3).}
\begin{tabular}{c c c c c c c c}
    \hline
   \multicolumn{2}{c}{}                                                   & \multicolumn{2}{c}{{$\nb \geq 0$}}        & \multicolumn{2}{c}{{$\nb = 0$}}             & \multicolumn{2}{c}{{$\nb \geq 1$}}      \\ \hline
                                            & \mll range (\GeVns)         & Pred.      & Obs.                              & Pred.  & Obs.                         & Pred.  & Obs.                        \\ \cline{2-8}

\multirow{10}{*}{Central}          & \multirow{2}{*}{20--70   }  &   477 $\pm$   30  &  { 445 } &   130 $\pm$   13  &  { 135 } &    347 $\pm$   24  &  { 310 } \\
&                                                                           & (   4.8 $\pm$    1.4) &                                    & (   3.6 $\pm$    1.1) &                                    &  (   1.2 $\pm$    0.3) &                                   \\ \cline{2-8}
                                            & \multirow{2}{*}{70--81 }  &   134 $\pm$   13  &  { 131 } &    40 $\pm$    6  &  {  33 } &     94 $\pm$   10  &  { 98 } \\
&                                                                           & (   2.7 $\pm$    0.8) &                                    & (   2.1 $\pm$    0.6) &                                    &  (   0.7 $\pm$    0.2) &                                   \\ \cline{2-8}
                                             & \multirow{2}{*}{81--101   } &   254 $\pm$   18  &  { 275 } &    95 $\pm$   11  &  { 107 } &    160 $\pm$   14  &  { 168 } \\
&                                                                           & (  62 $\pm$    8) &                                    & (  46 $\pm$    8) &                                    &  (  16 $\pm$    2) &                                   \\ \cline{2-8}
                                            & \multirow{2}{*}{101--120   }  &   166 $\pm$   15  &  { 165 } &    48 $\pm$    7  &  { 43 } &    118 $\pm$   12  &  { 122 } \\
&                                                                           & (   2.1 $\pm$    0.6) &                                    & (   1.6 $\pm$    0.5) &                                    &  (   0.5 $\pm$    0.2) &                                   \\ \cline{2-8}
                                             & \multirow{2}{*}{$>$120    } &   477 $\pm$   30  &  { 518 } &   112 $\pm$   12  &  { 144 } &    365 $\pm$   25  &  { 374 } \\
&                                                                           & (   1.6 $\pm$    0.5) &                                    & (   1.2 $\pm$    0.4) &                                    &  (   0.4 $\pm$    0.1) &                                   \\ \cline{2-8}
\hline
\multirow{10}{*}{Forward}          & \multirow{2}{*}{20--70   }  &   111 $\pm$   12  &  { 136 } &    36 $\pm$    6  &  {  45 } &     75 $\pm$    10  &  { 91 } \\
&                                                                           & (   1.6 $\pm$    0.4) &                                    & (   1.2 $\pm$    0.4) &                                    &  (   0.4 $\pm$    0.1) &                                   \\ \cline{2-8}
                                            & \multirow{2}{*}{70--81   }  &    47 $\pm$    7  &  {  50 } &    15 $\pm$    4  &  {  14 } &     32 $\pm$    6  &  { 36 } \\
&                                                                           & (   1.2 $\pm$    0.3) &                                    & (   0.9 $\pm$    0.3) &                                    &  (   0.3 $\pm$    0.1) &                                   \\ \cline{2-8}
                                            & \multirow{2}{*}{81--101   }  &    100 $\pm$   10  &  {  92 } &    45 $\pm$    6  &  {  39 } &     55 $\pm$    8  &  { 53 } \\
&                                                                           & (  24 $\pm$    3) &                                    & (  18 $\pm$    3) &                                    &  (   6.0 $\pm$    1.2) &                                   \\ \cline{2-8}
                                            & \multirow{2}{*}{101--120   }  &    78 $\pm$    10  &  {  51 } &    22 $\pm$    5  &  {  15 } &     55 $\pm$    8  &  { 36 } \\
&                                                                           & (   1.0 $\pm$    0.3) &                                    & (   0.7 $\pm$    0.2) &                                    &  (   0.2 $\pm$    0.1) &                                   \\ \cline{2-8}
                                            & \multirow{2}{*}{$>$120   }  &   308 $\pm$   25  &  { 306 } &    85 $\pm$   10  &  {  95 } &    223 $\pm$   20  &  { 211 } \\
&                                                                           & (   0.7 $\pm$    0.2) &                                    & (   0.5 $\pm$    0.2) &                                    &  (   0.2 $\pm$    0.1) &                                   \\ \cline{2-8}
\hline
\end{tabular}
\end{table}

The agreement between the observation and the prediction is found to be better than $1\,\sigma$ in most of the regions. The largest deviation found corresponds to
a local significance of $1.8\,\sigma$. This result is compatible with the null hypothesis provided the large number of signal regions.

\begin{figure}[!h]
\centering
    \includegraphics[width=0.6\textwidth]{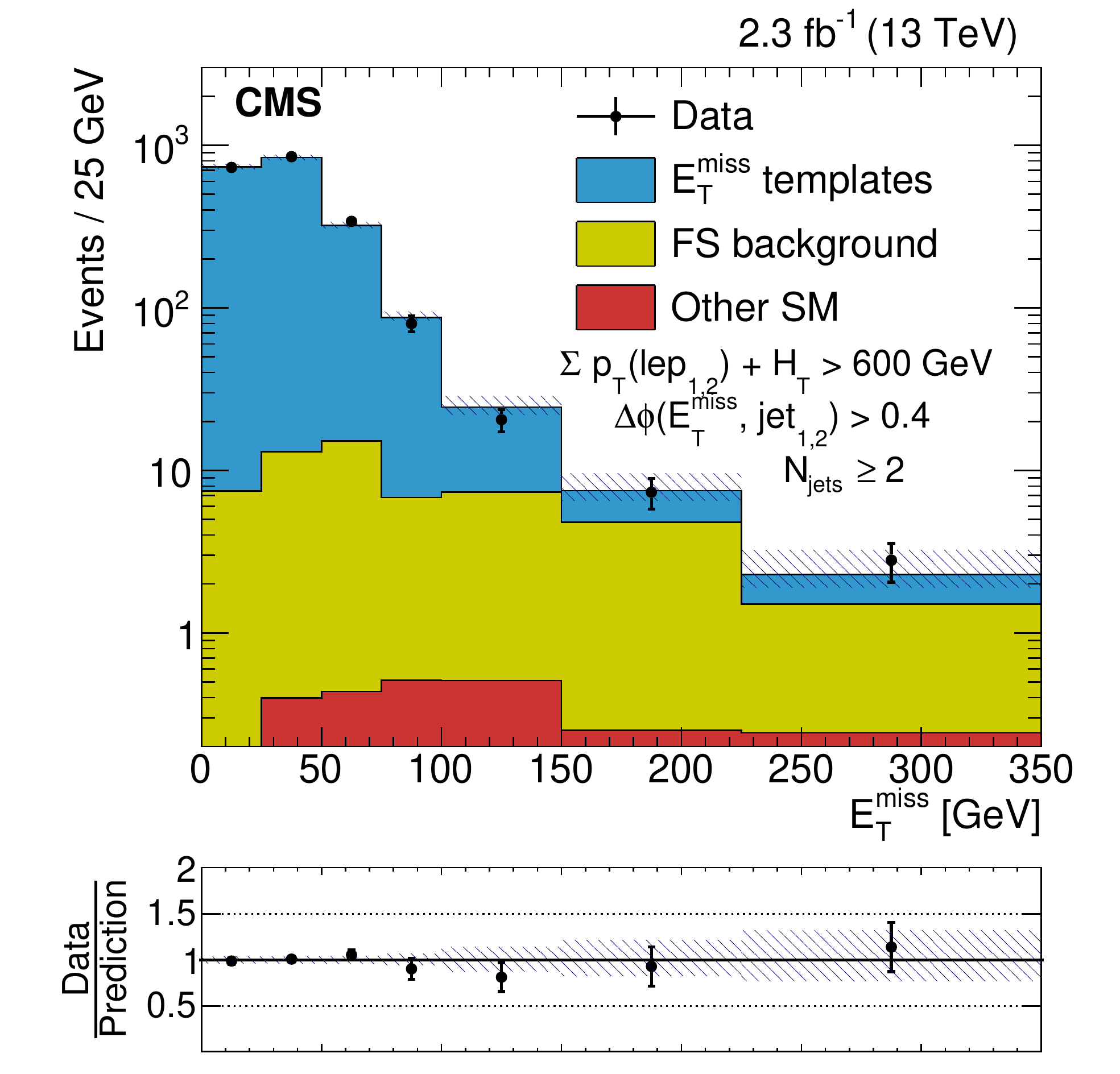}
    \includegraphics[width=0.6\textwidth]{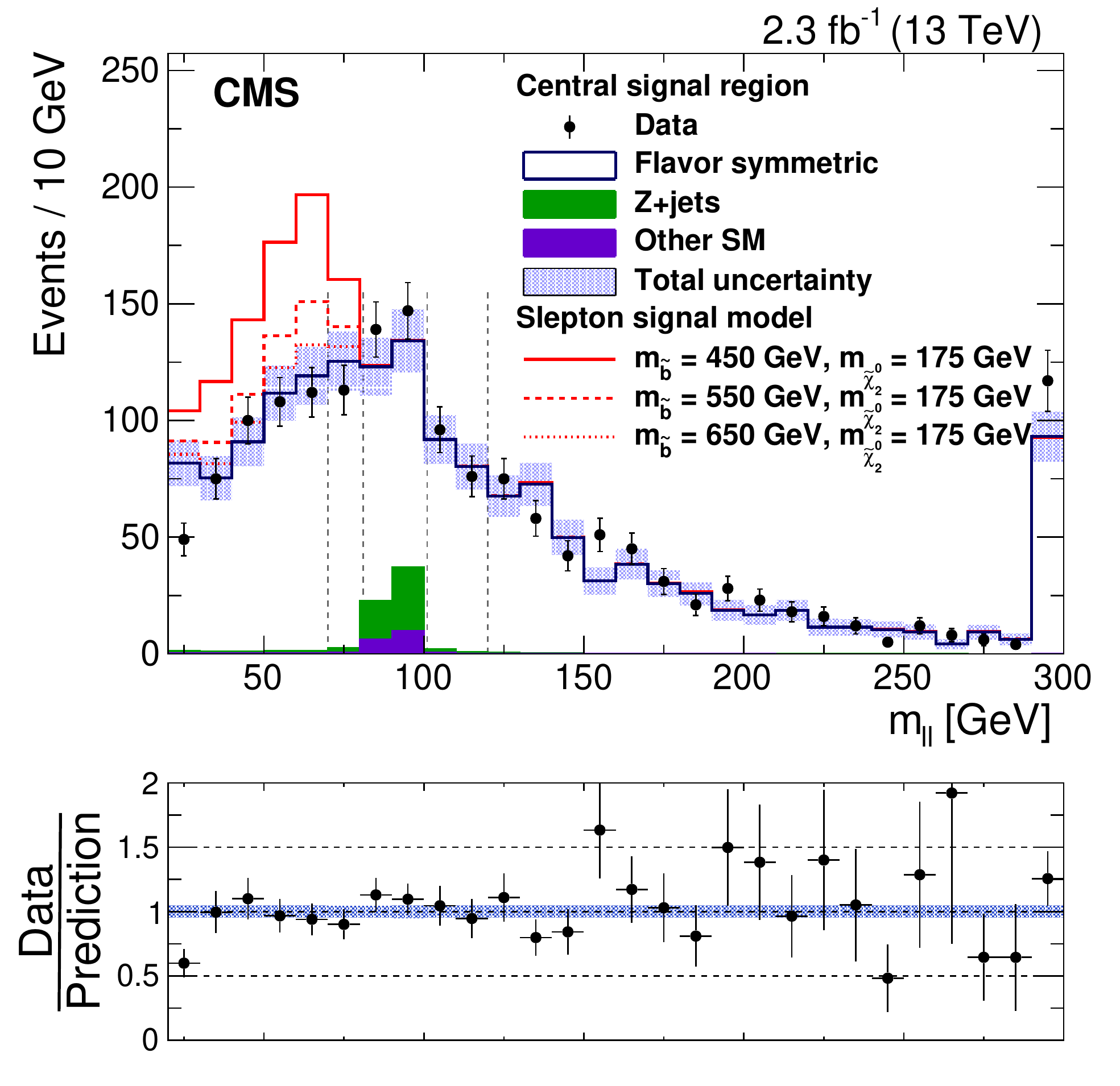}
    \caption{The \MET\ and \mll distributions are shown for data and background predictions in the on-Z ATLAS signal region (upper) and for the region where CMS reported an excess in Run 1 (lower). The ``Other SM'' category includes WZ, ZZ, and other rare SM backgrounds taken from MC. The red lines in the \mll distribution correspond to three different slepton-edge signal hypotheses overlaid on top of the background distribution.}
\label{fig:resultsOnZEdge}
\end{figure}

Figure~\ref{fig:resultsOnZEdge} (upper)
shows the \MET distribution for the on-Z ATLAS signal region, while Fig.~\ref{fig:resultsOnZEdge} (lower), shows the \mll distribution for the edge region
without any selection on the number of b-tagged jets and with central leptons, as in the region where CMS reported the excess at $\sqrt{s}=$ 8\TeV.
The comparison between the observation and prediction in these two regions of interest
does not indicate the presence of any excess with respect to the SM expectation. The $3.0\,\sigma$ discrepancy between observation and prediction in the first bin of
the \mll distribution in Fig.~\ref{fig:resultsOnZEdge} (lower), has been studied in detail in several control regions with similar kinematic properties, and also by modifying the trigger, identification and
isolation parameters of the leptons. Since no sign of any systematic effect has been found, we conclude this to be consistent with a statistical fluctuation.

\section{Interpretation}
\label{sec:interpretation}

The results of the analysis are interpreted in terms of simplified models. In order to quantify
the sensitivity of the on-Z and edge searches, two simulated samples with a scan of mass points of the GMSB and slepton-edge models have
been produced. Upper limits on the cross section multiplied by the branching ratio have been calculated at a 95\% confidence level (CL) using the CL$_S$ criterion and an asymptotic formulation~\cite{Junk:1999kv,0954-3899-28-10-313,HiggsTool1,Cowan:2010js}, taking into
account the statistical and systematic uncertainties in the signal yields and the background predictions.

\subsection{Systematic uncertainty in the signal yield}

The systematic uncertainties in the signal yield have been evaluated by comparing the yields obtained after making a variation
on the source of the systematic effect and the nominal yields. The uncertainty related to the measurement of the integrated luminosity is 2.7\%~\cite{CMS-PAS-LUM-15-001}.
The uncertainty in the corrections used to account for lepton identification and isolation efficiency differences between data and simulation is 2--4\% in the signal acceptance.
The uncertainty in the b tagging efficiency and mistag probability are 2--5\% except for the edge signal regions without b tags, where they can range up to 20\%. A further
systematic uncertainty of 1--6\% is considered on the scale factors correcting for the differences between fast and \GEANTfour simulations for leptons. Dilepton trigger efficiencies ranging
between 87\% and 96\%, and depending on the lepton flavor, are measured in data and
applied as an overall scale factor to the signal simulation with a systematic uncertainty of 5\%. The uncertainty in the jet energy scale varies between
0\% and 8\% depending on the signal kinematics. The uncertainty associated with the modeling of initial-state radiation (ISR) is 1--3\%.
The uncertainty in the correction to account for the pileup in the simulation is evaluated by shifting the inelastic cross section by $\pm$5\%
and amounts to less than 6\% on signal acceptance. Finally the statistical uncertainty on the number of simulated events is also considered and found
to be in the range 1--20\%, where the regions with low population of signal due to the acceptance in \MET and/or b-tag multiplicity are most affected. These uncertainties are summarized in Table~\ref{tab:systs}.

\begin{table}[htb]
\renewcommand{\arraystretch}{1.2}
\centering
\topcaption{\label{tab:systs}
List of systematic uncertainties taken into account for the signal yields and typical values.}
\begin{tabular}{l c}
\hline
Source of uncertainty                         & Uncertainty (\%)     \\ \hline
Luminosity                                    & 2.7                  \\
Pileup                                        & 0--6              	 \\
b tag modeling                                & 2--20                 \\
Lepton reconstruction and isolation           & 2--4                    \\
Fast simulation scale factors    	      & 1--6                  \\
Trigger modeling                              & 5                    \\
Jet energy scale                              & 0--8                  \\
ISR  modeling                                 & 1--3                  \\
Statistical uncertainty                       & 1--20                 \\ \hline
Total uncertainty                    & 7--32                 \\
\hline
\end{tabular}
\end{table}

\subsection{Interpretation using simplified models}

\begin{figure}[!hb]
\centering
\includegraphics[width=0.6\textwidth]{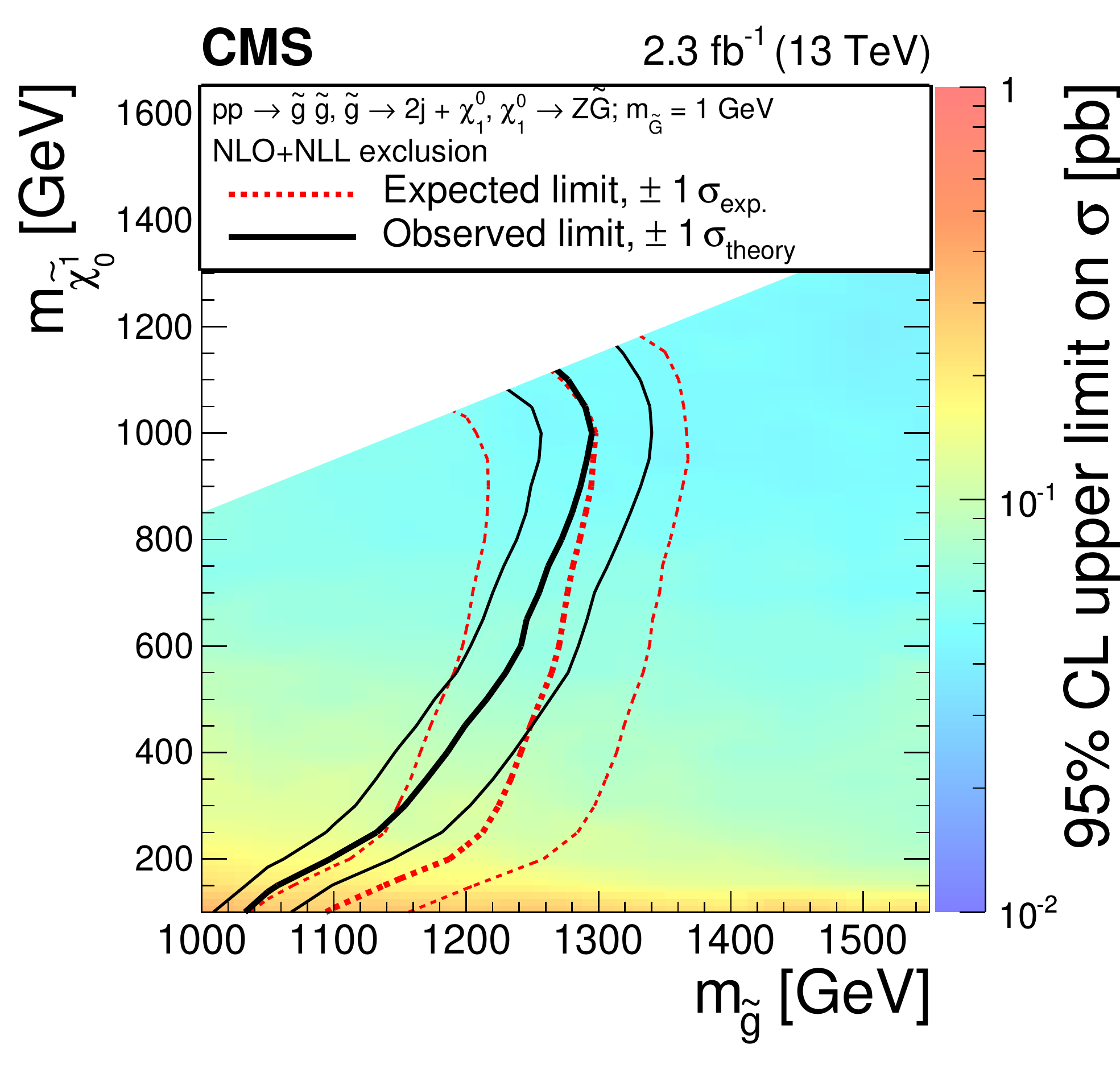}
\caption{Cross section upper limits and exclusions contours at 95\% CL with the results of the on-Z search interpreted in the GMSB model. The region to the left of the red dotted (black solid) line shows the masses which are excluded by the expected (observed) limit.}
\label{fig:Limits1}
\end{figure}

Since the GMSB model leads to a signature containing at least 6 jets in the final state, most of the sensitivity of the
on-Z search is provided by the high jet multiplicity signal regions defined within the SRB category. We only consider the number of
observed and predicted events in these regions to set limits on this model. The expected and observed limits are presented in Fig.~\ref{fig:Limits1}. We exclude
gluino masses up to 1.28\,(1.03)\TeV for large (small) neutralino masses. These results show an improvement with respect to the 8\TeV result where
we obtained an observed and expected limits for gluino masses from 1.0 to 1.1\TeV.

\begin{figure}[!hb]
\centering
\includegraphics[width=0.6\textwidth]{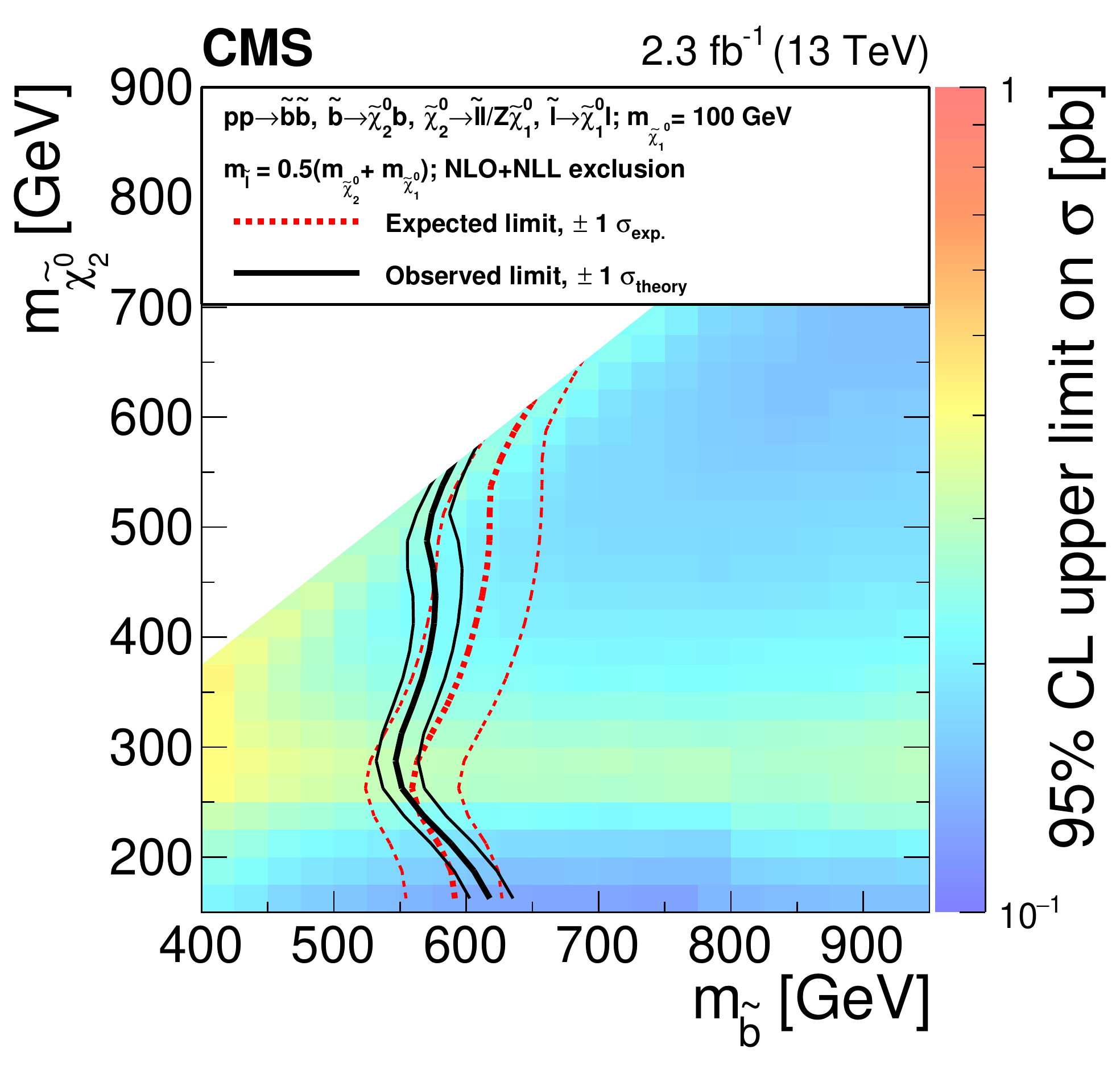}
\caption{Cross section upper limits and exclusion contours at 95\% CL with the results of the edge search interpreted in the slepton-edge model. The region to the left
of the red dotted (black solid) line shows the masses which are excluded by the expected (observed) limit.}
\label{fig:Limits2}
\end{figure}

The edge search is interpreted using the slepton-edge model, combining all the invariant mass, $\abs{\eta}$, and mutually exclusive b tag regions. Figure~\ref{fig:Limits2} shows
the exclusion contour in the plane of the masses of the bottom squark and the second neutralino. We exclude bottom squark masses up to 620\GeV at low \PSGczDt masses.
The slight decrease in sensitivity at a neutralino mass of $\sim$250\GeV corresponds to a kinematic edge located at $\sim$150--200\GeV. In this case the signal is spread evenly
across all mass regions, while in the case of low (high) \PSGczDt masses, the majority of signal events fall into the low- (high-) mass bin, which increases the sensitivity for these mass points.
The expected upper limits in the bottom squark/neutralino mass plane are similar to the limits set by the 8\TeV analysis.
In two parameter regions the expected limits are slightly improved due to the introduction of new signal regions. The introduction of the below-Z and above-Z signal
region increases the  sensitivity of the analysis for sbottom masses of about 550\GeV and neutralino masses of around 250\GeV. The second improvement is the categorization
according to the number of b-tagged jets that gives additional sensitivity close to the sbottom and neutralino mass diagonal where events with zero b-tagged jets become important since the produced b jets have less energy and are often not identified. The observed upper limits in the region with small neutralino masses have been largely improved with respect to the 8\TeV results from 500 to approximately 620\GeV.

\section{Summary}
\label{sec:summary}

A search for physics beyond the standard model has been presented in the opposite-sign, same-flavor lepton final state using a data sample of
pp collisions collected at a center-of-mass energy of 13\TeV, corresponding to an integrated luminosity of \lint\fbinv, recorded with
the CMS detector in 2015. Searches are performed for signals that either produce a kinematic edge, or a peak at the Z boson mass, in the dilepton
invariant mass distribution. Comparing the observation to estimates for SM backgrounds obtained from data control samples, no statistically significant
evidence for a signal has been observed. Notably, this is true for the two event selections where excesses of 2.6 and $3.0\,\sigma$ significance had
been observed by the CMS and ATLAS collaborations in their respective 8\TeV results~\cite{CMS:edge,ATLAS:edge}.

The search for events containing an on-shell Z boson is interpreted in a model of gauge-mediated supersymmetry breaking, where
the Z bosons are produced in decay chains initiated through gluino pair production, and where the branching ratios have been fixed to 100\% to produce the desired topology. Gluino masses below 1.28\TeV for high neutralino masses
and 1.03\TeV for low neutralino masses have been excluded, extending the previous exclusion limits derived from a similar analysis at 8\TeV by almost 200\GeV.

The search for an edge is interpreted in a simplified model based on bottom squark pair production, where dilepton mass edges are produced in decay chains containing
the two lightest neutralinos and a slepton, where again the branching ratios have been fixed to produce the desired topology. Bottom squark masses below 550 and 620\GeV
have been excluded, depending on the \PSGczDt mass. These limits are similar to previous exclusion limits except for low \PSGczDt masses where the excluded limits have been extended by about 100\GeV.

\begin{acknowledgments}
We congratulate our colleagues in the CERN accelerator departments for the excellent performance of the LHC and thank the technical and administrative staffs at CERN and at other CMS institutes for their contributions to the success of the CMS effort. In addition, we gratefully acknowledge the computing centers and personnel of the Worldwide LHC Computing Grid for delivering so effectively the computing infrastructure essential to our analyses. Finally, we acknowledge the enduring support for the construction and operation of the LHC and the CMS detector provided by the following funding agencies: BMWFW and FWF (Austria); FNRS and FWO (Belgium); CNPq, CAPES, FAPERJ, and FAPESP (Brazil); MES (Bulgaria); CERN; CAS, MoST, and NSFC (China); COLCIENCIAS (Colombia); MSES and CSF (Croatia); RPF (Cyprus); SENESCYT (Ecuador); MoER, ERC IUT and ERDF (Estonia); Academy of Finland, MEC, and HIP (Finland); CEA and CNRS/IN2P3 (France); BMBF, DFG, and HGF (Germany); GSRT (Greece); OTKA and NIH (Hungary); DAE and DST (India); IPM (Iran); SFI (Ireland); INFN (Italy); MSIP and NRF (Republic of Korea); LAS (Lithuania); MOE and UM (Malaysia); BUAP, CINVESTAV, CONACYT, LNS, SEP, and UASLP-FAI (Mexico); MBIE (New Zealand); PAEC (Pakistan); MSHE and NSC (Poland); FCT (Portugal); JINR (Dubna); MON, RosAtom, RAS and RFBR (Russia); MESTD (Serbia); SEIDI and CPAN (Spain); Swiss Funding Agencies (Switzerland); MST (Taipei); ThEPCenter, IPST, STAR and NSTDA (Thailand); TUBITAK and TAEK (Turkey); NASU and SFFR (Ukraine); STFC (United Kingdom); DOE and NSF (USA).

Individuals have received support from the Marie-Curie program and the European Research Council and EPLANET (European Union); the Leventis Foundation; the A. P. Sloan Foundation; the Alexander von Humboldt Foundation; the Belgian Federal Science Policy Office; the Fonds pour la Formation \`a la Recherche dans l'Industrie et dans l'Agriculture (FRIA-Belgium); the Agentschap voor Innovatie door Wetenschap en Technologie (IWT-Belgium); the Ministry of Education, Youth and Sports (MEYS) of the Czech Republic; the Council of Science and Industrial Research, India; the HOMING PLUS program of the Foundation for Polish Science, cofinanced from European Union, Regional Development Fund, the Mobility Plus program of the Ministry of Science and Higher Education, the OPUS program contract 2014/13/B/ST2/02543 and contract Sonata-bis DEC-2012/07/E/ST2/01406 of the National Science Center (Poland); the Thalis and Aristeia programs cofinanced by EU-ESF and the Greek NSRF; the National Priorities Research Program by Qatar National Research Fund; the Programa Clar\'in-COFUND del Principado de Asturias; the Rachadapisek Sompot Fund for Postdoctoral Fellowship, Chulalongkorn University and the Chulalongkorn Academic into Its 2nd Century Project Advancement Project (Thailand); and the Welch Foundation, contract C-1845.
\end{acknowledgments}

\bibliography{auto_generated}

\cleardoublepage \appendix\section{The CMS Collaboration \label{app:collab}}\begin{sloppypar}\hyphenpenalty=5000\widowpenalty=500\clubpenalty=5000\textbf{Yerevan Physics Institute,  Yerevan,  Armenia}\\*[0pt]
V.~Khachatryan, A.M.~Sirunyan, A.~Tumasyan
\vskip\cmsinstskip
\textbf{Institut f\"{u}r Hochenergiephysik der OeAW,  Wien,  Austria}\\*[0pt]
W.~Adam, E.~Asilar, T.~Bergauer, J.~Brandstetter, E.~Brondolin, M.~Dragicevic, J.~Er\"{o}, M.~Flechl, M.~Friedl, R.~Fr\"{u}hwirth\cmsAuthorMark{1}, V.M.~Ghete, C.~Hartl, N.~H\"{o}rmann, J.~Hrubec, M.~Jeitler\cmsAuthorMark{1}, A.~K\"{o}nig, I.~Kr\"{a}tschmer, D.~Liko, T.~Matsushita, I.~Mikulec, D.~Rabady, N.~Rad, B.~Rahbaran, H.~Rohringer, J.~Schieck\cmsAuthorMark{1}, J.~Strauss, W.~Treberer-Treberspurg, W.~Waltenberger, C.-E.~Wulz\cmsAuthorMark{1}
% ...
\vskip\cmsinstskip
% ...
1:~~Also at Vienna University of Technology, Vienna, Austria\\
% ...

\end{sloppypar}
\end{document}